\documentclass{PoS}

\makeatletter

\newcommand{\gsim}
{\;\raisebox{-.3em}{$\stackrel{\displaystyle >}{\sim}$}\;}

\def\dsl{\mathpalette\make@slash}
\def\make@slash#1#2{\setbox\z@\hbox{$#1#2$}%
  \hbox to 0pt{\hss$#1/$\hss\kern-\wd0}\box0}

\makeatother

\def\reffi#1{\mbox{Fig.~\ref{#1}}}

\def\citere#1{\mbox{Ref.~\cite{#1}}}

\def\mathswitch#1{\relax\ifmmode#1\else$#1$\fi}
\def\mathswitchr#1{\relax\ifmmode{\mathrm{#1}}\else$\mathrm{#1}$\fi}
\def\mathswitchit#1{\relax\ifmmode{#1}\else$#1$\fi}

\newcommand{\PW}{\mathswitchr W}
\newcommand{\PZ}{\mathswitchr Z}

\newcommand{\PH}{\mathswitchr H}

\newcommand{\Pl}{\mathswitch l}

\newcommand{\Pb}{\mathswitchr b}

\newcommand{\Pp}{\mathswitchr p}

\newcommand{\MH}{\mathswitch {M_\PH}}

\newcommand{\GeV}{\unskip\,\mathrm{GeV}}

\newcommand{\TeV}{\unskip\,\mathrm{TeV}}

\title{EW corrections to Higgs strahlung \\ 
at the Tevatron and the LHC with HAWK%
}

\ShortTitle{EW corrections to Higgs strahlung with HAWK}

\author{Ansgar Denner \\ 
	Universit\"at W\"urzburg, Institut f\"ur Theoretische Physik und Astrophysik\\ D-97074 W\"urzburg, Germany \\
        E-mail: \email{denner@physik.uni-wuerzburg.de}}

\author{\speaker{Stefan Dittmaier} \\
        Albert-Ludwigs-Universit\"at Freiburg, Physikalisches Institut\\ D-79104 Freiburg, Germany \\
        E-mail: \email{stefan.dittmaier@physik.uni-freiburg.de}}

\author{Stefan Kallweit \\
        Paul Scherrer Institut, W\"urenlingen und Villigen\\ CH-5232 Villigen PSI, Switzerland \\
	and\\
	Universit\"at Z\"urich, Institut f\"ur Theoretische Physik\\
	CH-8057 Z\"urich, Switzerland \\
        E-mail: \email{kallweit@physik.uzh.ch}}

\author{Alexander M\"uck \\
        RWTH Aachen University, Institut f\"ur Theoretische Teilchenphysik 
	und Kosmologie \\ D-52056 Aachen, Germany \\
        E-mail: \email{mueck@physik.rwth-aachen.de}}

\abstract{We briefly report on the inclusion of NLO QCD and electroweak
corrections to the Higgs-strahlung processes
$\Pp\Pp/\Pp\bar\Pp\to\PH\PW/\PZ\to\PH+2\,$leptons in the
Monte Carlo program {\sc Hawk}\footnote{The program is publically available under
http://omnibus.uni-freiburg.de/sd565/programs/hawk/hawk.html .}.
The electroweak corrections, which are
at the level of $-(5{-}10)\%$ for total cross sections, further increase
in size with increasing transverse momenta ($p_{\mathrm{T}}$) in differential cross
sections. For instance, for $p_{\mathrm{T,H}}\gsim200\GeV$, which
is the interesting range at the LHC, the
electroweak corrections to WH production
reach about $-15\%$ for $\MH=120\GeV$.
          }

\FullConference{The 2011 Europhysics Conference on High Energy Physics, EPS-HEP 2011,\\
		July 21-27, 2011\\
		Grenoble, Rh\^one-Alpes, France}

\begin{document}

\section{Introduction}

The Higgs-strahlung processes 
$\Pp\Pp/\Pp\bar\Pp\to\PH\PW/\PZ\to\PH+2\,$leptons
represent important search channels for a Standard Model 
Higgs boson in the low-mass range, both at the Tevatron and the LHC.
The challenge in the corresponding experimental 
analyses is background control, in particular at the LHC where
the Higgs boson is reconstructed from the jet substructure in the
decay $\PH\to\Pb\bar\Pb$ at high transverse momenta~\cite{Butterworth:2008iy}.

On the theoretical side, at least next-to-leading-order (NLO) corrections
should be taken into account to match the required precision in the analyses,
keeping as much as possible the differential information.
The QCD corrections can be classified into two different categories:
the Drell--Yan-like contributions, which respect factorization
according to $\Pp\Pp\to V^*\to\PH V$ and comprise the dominant 
correction, and a remainder, which contributes at 
next-to-next-to-leading-order (NNLO) first.
The former are known to NNLO for the total cross section~\cite{Brein:2003wg}
and for WH production also for differential distributions~\cite{Ferrera:2011bk}; 
the latter have been calculated at NNLO for total cross sections recently~\cite{nnloqcdrem}.
Electroweak (EW) corrections were first evaluated for the total cross sections
in \citere{Ciccolini:2003jy} and turn out to be about $-(5{-}10)\%$.
An update of the cross-section prediction for the LHC, 
together with a thorough estimate
of uncertainties, was recently presented in the report~\cite{Dittmaier:2011ti}
of the LHC Higgs Cross Section Working Group.
For the LHC with a centre-of-mass (CM) energy of $7(14)\TeV$
the QCD scale uncertainties
were assessed to be about $1\%$ and $1{-2}(3{-}4)\%$ for $\PW\PH$
and $\PZ\PH$ production, respectively, while
uncertainties of the parton distribution functions (PDFs) turn out
to be about $3{-}4\%$.

Here we briefly report on results of the first evaluation~\cite{whzhhawk} 
of EW NLO corrections to the full processes
$\Pp\Pp/\Pp\bar\Pp\to\PH\PW/\PZ\to\PH+2\,$leptons, supporting the
complete kinematical information of the decay products.
The corresponding calculation is included in the Monte Carlo program {\sc Hawk},
which was originally designed for the description of Higgs
production via vector-boson fusion including NLO QCD and EW
corrections~\cite{Ciccolini:2007jr}.

\section{Numerical results on NLO corrections}

In the following we exemplarily show results on the transverse-momentum 
($p_{\mathrm{T}}$)
distributions of the Higgs boson for the various leptonic decays of the
W/Z bosons at the Tevatron and the LHC with a CM energy of
$7\TeV$, using the input parameters of the LHC Higgs Cross Section 
Group~\cite{Dittmaier:2011ti} and the basic identification cuts
$p_{\mathrm{T},l}>20\GeV$, $|y_l|<2.5$, $\dsl{p}_{\mathrm{T}}>25\GeV$.
For the LHC we additionally require 
$p_{\mathrm{T,H}}>200\GeV$ and $p_{\mathrm{T,W/Z}}>190\GeV$ to
mimic the analysis that demands a highly-boosted Higgs boson. The slightly 
asymmetric cuts are chosen to avoid large NLO corrections in the $p_{\mathrm{T,H}}$
distribution near the cut which would appear in fixed-order 
calculations for symmetric cuts. 

\begin{figure}
\includegraphics[width=7.5cm]{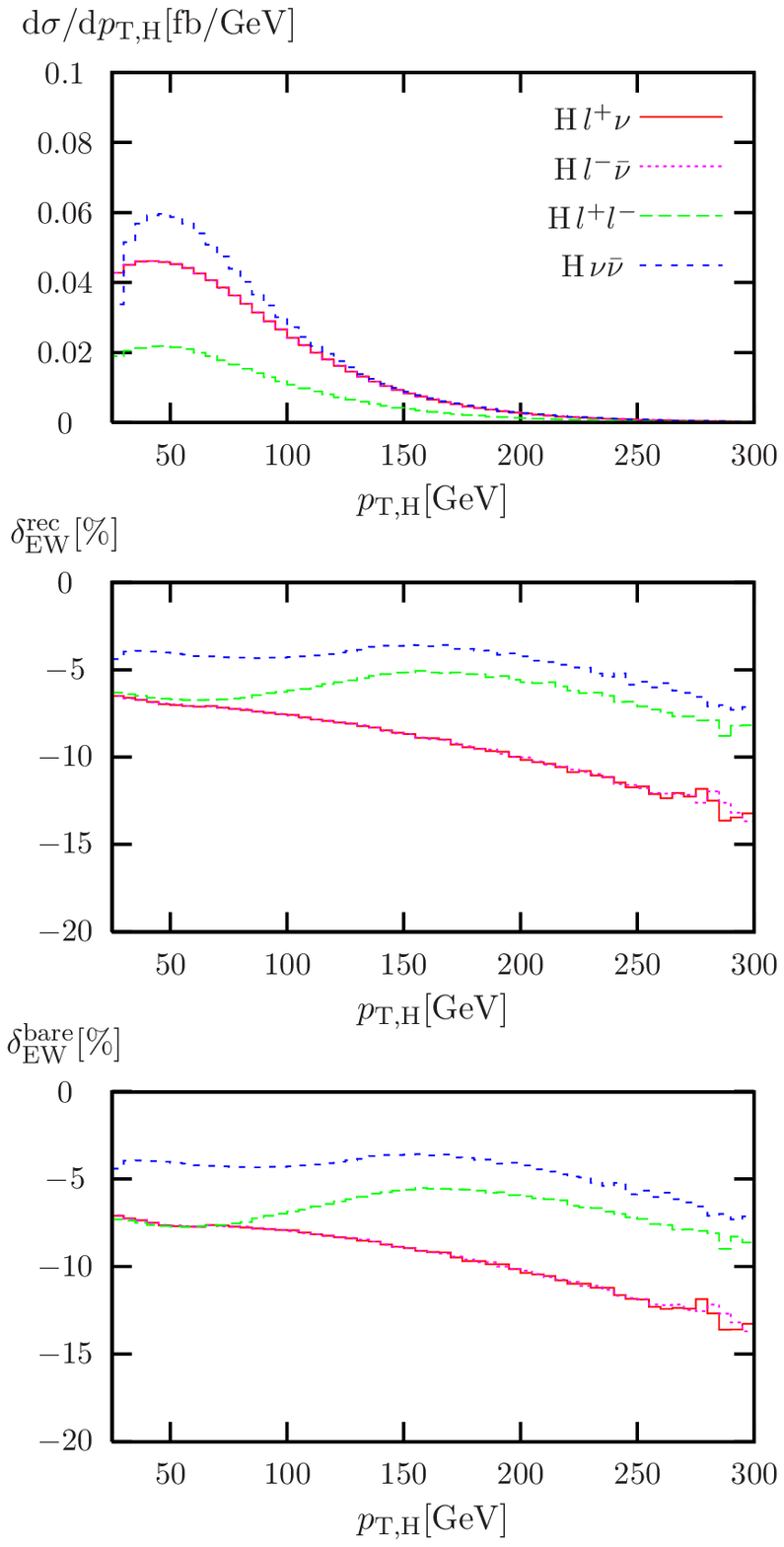}
\includegraphics[width=7.5cm]{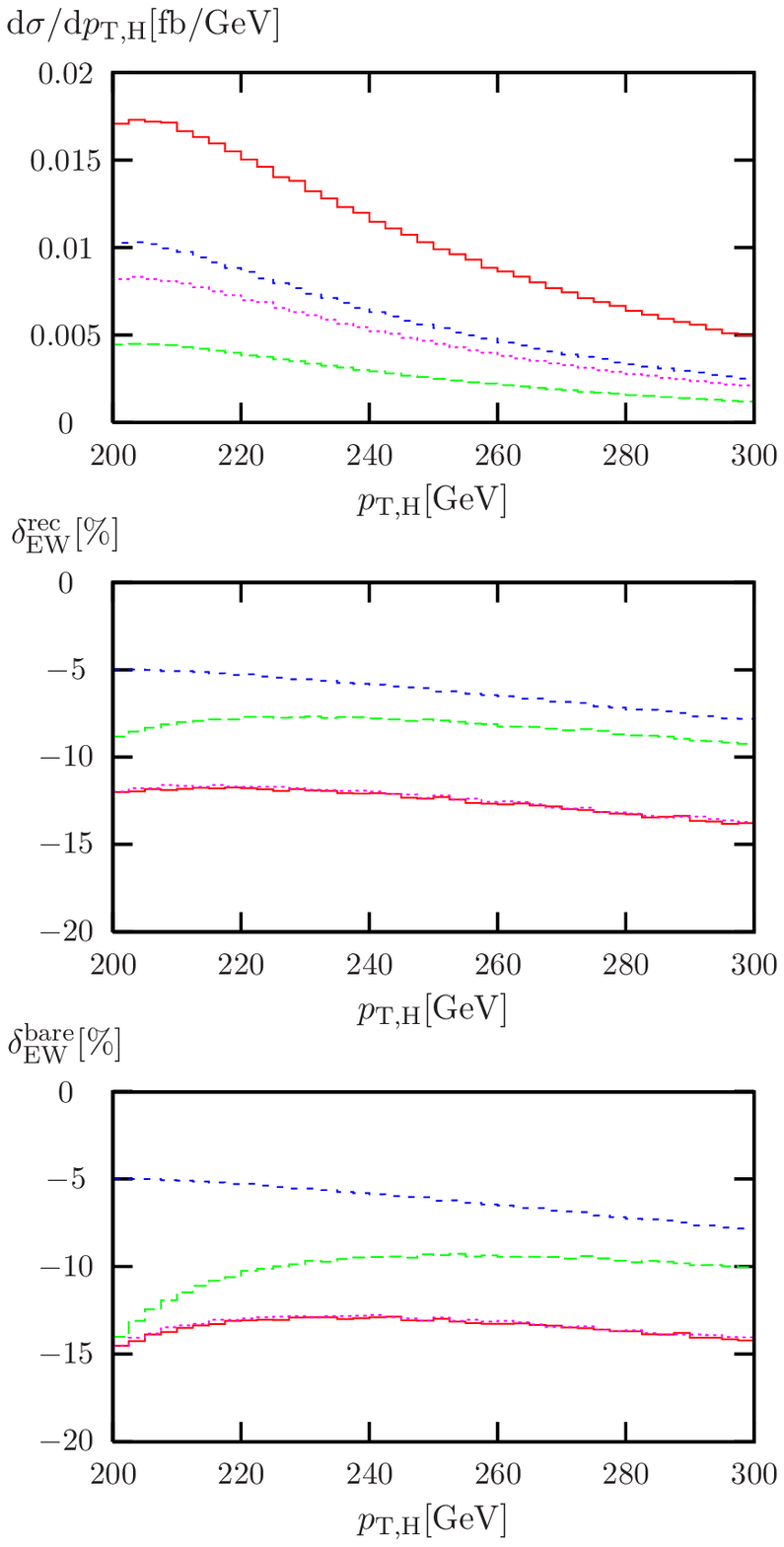}
\caption{\label{fi:pth}
{\sc Hawk} prediction for the absolute $p_{\mathrm{T},\PH}$ distributions
(top) and NLO EW corrections for recombined (middle)
and bare (bottom) leptons for Higgs strahlung
with basic cuts at the Tevatron (left) as well as for boosted Higgs bosons at 
the 7 TeV LHC (right) for $\MH=120$ GeV.
} 
\end{figure}
\begin{figure}
\begin{center}
\includegraphics[width=15cm]{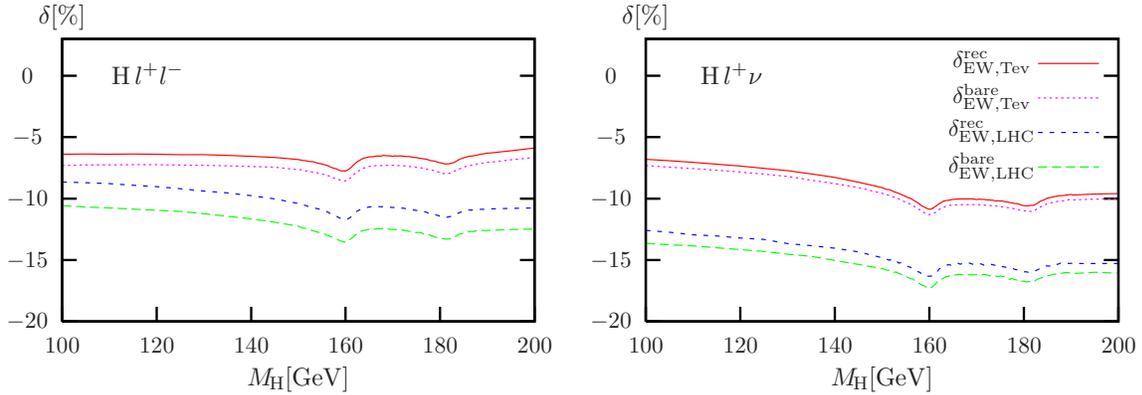}
\end{center}
\caption{\label{fig:mH}
{\sc Hawk} prediction for the EW corrections to the cross sections 
for $\PH \Pl^+\Pl^-$  (left) and $\PH \Pl^+\nu_\Pl$ (right) production
with basic cuts at the Tevatron and for boosted Higgs bosons at 
the 7 TeV LHC as a function of the Higgs-boson mass.}
\end{figure}

Figure~\ref{fi:pth} shows the differential cross sections
with respect to the transverse momentum
of the Higgs boson up to $300\GeV$, along with the EW corrections, and 
Figure~\ref{fig:mH} contains the corresponding integrated quantities
as a function of the Higgs-boson mass. All results are given for a specific 
leptonic decay mode and are not summed over lepton generations.
The two different versions ``rec'' and ``bare'' of the EW corrections
refer to different treatments of final-state photons. The latter
assumes perfect isolation of photons that are collinear to the lepton,
which is then assumed to be a muon, the former recombines photons
and leptons that are very close 
($R_{\gamma l} < 0.1$), which is closer to the treatment of electrons.
The results for invisible \PZ\ decays, 
of course, do not depend on the neutrino flavour and can be trivially obtained 
by multiplying the $\PH \nu_\Pl \overline{\nu}_\Pl$ results by three.
For the Higgs-boson sample at the Tevatron and the inclusive sample
at the LHC including low $p_\mathrm{T,\PH}$ (not shown here),
the EW corrections range between $-5\%$ and $-15\%$ and are largest for
the \PW-mediated channels at large $p_\mathrm{T,\PH}$.
For increasing $p_\mathrm{T,\PH}$ the EW corrections become more and more
pronounced and show the onset of the large EW logarithms from
soft/collinear W/Z exchange at high energies.
The difference due to the treatment
of final-state photons is small, since the bulk of the correction 
is of weak origin and not due to final-state radiation. 
The same observation holds for large values of the transverse momentum in the
boosted-Higgs analysis at the LHC. Only for $p_\mathrm{T,\PH}$ close to 200~GeV,
the influence of final-state radiation is larger, because it may shift the
measured value of $p_{\mathrm{T},V}$, being equal to $p_\mathrm{T,\PH}$
without radiation, below the corresponding cut value.
The EW corrections for the corresponding integrated
cross sections (see \reffi{fig:mH}) are again significantly larger 
for the LHC setup compared to the inclusive sample which is dominated by
events with small $p_\mathrm{T,\PH}$. For the \PW\PH\
channel they reach $-15\%$. There are additional EW corrections at the level of +1\% (not included in 
the figures) that are due to photons in the initial state and require a photon 
distribution function.

Using the complex-mass scheme to consistently treat resonant vector bosons in the
calculation, also the threshold regions ($M_\PH\sim 2 M_\PW$ and
$M_\PH\sim 2 M_\PZ$), which were not properly described
in the past~\cite{Ciccolini:2003jy}, are under theoretical control.

A wider selection of results and a description of the calculational details
can be found in \citere{whzhhawk}.

\end{document}